\documentclass[a4paper,11pt]{article}

\usepackage{amsmath,amssymb}    
\usepackage{color}
\usepackage{graphicx}
\usepackage{hyperref}            
\usepackage{multirow,makecell}   
\usepackage{float}
\usepackage[utf8]{inputenc}
\numberwithin{equation}{section}

\def \nn {\nonumber}
\DeclareFixedFont\trfont{OT1}{phv}{b}{sc}{11}

\begin{document}

\thispagestyle{empty}
\begin{flushright}\footnotesize
	\texttt{NORDITA-2019-015} \\
	
\end{flushright}

\renewcommand{\thefootnote}{\fnsymbol{footnote}}
\setcounter{footnote}{0}

\begin{center}
	{\Large\textbf{\mathversion{bold} 
			Holographic four-point functions in Toda field theories in $AdS_2$   
		}
		\par}
	
	\vspace{0.3cm}
	
	\textrm{Hao Ouyang$^{1}$}
	\vspace{4mm}
	
	{\small 
		\textit{${}^1$Nordita, KTH Royal Institute of Technology and Stockholm University\\
			Roslagstullsbacken 23, SE-106 91 Stockholm, Sweden}\\
				\texttt{hao.ouyang@su.se}
	}

	

	\par\vspace{0.5cm}
	
	\textbf{Abstract} \vspace{3mm}
	
	\begin{minipage}{\textwidth}
		We consider Toda field theories in a classical Euclidean $AdS_2$ background.
		 We compute the four-point functions of  boundary operators in the $a_1$, $a_2$ and $b_2$ Toda field theories. 
		 They take the same form as  the  four-point functions of generators in the corresponding $\mathcal{W}$-algebras.
		Therefore we conjecture that the boundary operators are in one-to-one correspondence
		with the generators in the $\mathcal{W}$-algebras.
	\end{minipage}
	
\end{center}

\vspace{1.5cm}

\section{Introduction}
Among two-dimensional field theories, integrable field theories distinguish themselves by the existence of an infinite number of conserved
 charges. 
In flat spacetime, two basic consequences of higher conservation laws are
 the absence of particle production and factorization of the higher-point S-matrix\cite{Zamolodchikov:1978xm,Dorey:1996gd,Shankar:1977cm,Parke:1980ki}.
Since  a Lax pair is in general hard to find, 
no tree-level particle production is regarded as an important sign of classical integrability, 
and it is a useful strategy for building classically integrable theories \cite{Arefeva:1974bk,Braden:1991vz,Dorey:1996gd,Gabai:2018tmm,Bercini:2018ysh}.

It is natural to wonder  what one can say about integrable field theories in other two-dimensional spacetimes. The $AdS_2$ case is of particular interest since it is maximally symmetric and has applications related to the AdS/CFT correspondence\cite{Maldacena:1997re,Witten:1998qj,Gubser:1998bc}.
S-matrix cannot be defined in AdS spacetime due to the periodicity of particle orbits and the timelike boundary.
The AdS analogs of the flat space S-matrix elements  are  boundary correlation functions which can be obtained as a limit of bulk correlation functions.
In the weak coupling regime, they can be computed by Witten diagrams. Such observables  in $AdS_2$ can be used to define a set of correlation functions in a one-dimensional conformal theory \cite{Paulos:2016fap}.
Due to the existence of infinitely many conservation laws, it is intuitive to expect  there should be something special about boundary correlation functions in integrable field theories.

Integrable field theories in $AdS_2$ are rare.
Toda field theories with simple Lie algebras provide an important class of such examples.
Toda field theories are conformal \cite{Gervais:1983am,Mansfield:1982sq,Braaten:1983pz} and  integrable\cite{Leznov:1979td,mikhailov1979integrability,Olive:1983mw,Olive:1984mb}.
It was shown in \cite{Bilal:1988ze,Bilal:1988jf,Bilal:1988jg,Bilal:1988nf} that they possess extended symmetries generated by  $\mathcal{W}$-algebras \cite{Zamolodchikov:1985wn} which are higher spin extensions of the Virasoro algebra (see \cite{Bouwknegt:1992wg} for a review of $\mathcal{W}$-algebras).

In this paper we focus on four-point functions of  boundary operators in Toda field theories in a classical $AdS_2$ background.
Unlike previous efforts devoted to the correlation functions of the vertex operators,
 we are interested in the boundary operators obtained by pushing bulk scalar fields to the conformal boundary.
 The conformal dimensions of the boundary operators are related to the masses of the bulk fields.
The potentials of Toda field theories in flat space  do not have a stable minimum.
When a Toda field theory is put in an arbitrary Riemann surface, the scalar fields need to couple to curvature  in order to preserve integrability.
Since the $AdS_2$ spacetime or its Euclidean version has a negative constant scalar curvature, 
the potential acquires a  minimum and the scalar fields become massive. The spectrum of the theory is determined by the set of exponents of the associated Lie algebra.
We compute the tree-level four-point functions in  Toda field theories with the Lie algebras $a_1$, $a_2$ and $b_2$.
Adding up contributions from the exchange Witten diagrams and contact Witten diagrams, the final expressions are rather simple.
Indeed, they are related to the  four-point functions of generators in $\mathcal{W}$-algebra in the limit of large central charge.

This paper is organized as follows. In section \ref{s2} we briefly review Toda field theories in $AdS_2$. 
In section \ref{s3}, we present in detail the calculation of four-point functions in the $a_1$, $a_2$ and $b_2$ Toda field theories.
Conclusions are given in section \ref{sc}.
Some reduced $\bar D$-functions used in the calculation are collected in appendix \ref{app1}.

\section{Toda field theories in $AdS_2$}\label{s2}
The action of a Toda field theory associated with a finite-dimensional simple Lie algebras $\mathfrak{g}$ of rank $r$ in an Euclidean $AdS_2$ (hyperbolic plane) is given by
\begin{equation}
S_{\mathfrak{g}}=\int d^2x  \sqrt{g}  \Big(
\frac{1}{2}g^{\mu \nu}\partial_\mu \phi \cdot \partial_\nu \phi+V_{\mathfrak{g}}(\phi)
\Big),
\end{equation}
with
\begin{equation}
V_{\mathfrak{g}}(\phi)= \frac{1}{\beta^2}\sum_i^r n_i e^{\beta \alpha_i \cdot \phi}+\frac{1}{2}R Q \cdot\phi,
\end{equation}
where $\alpha_i$ are simple roots of $\mathfrak{g}$ and  $\phi$ is an $r$-component scalar field.
We consider an $AdS_2$ with unit radius. The metric in Poincar\'e coordinates is
\begin{equation}
ds^2=\frac{dx^2+dz^2}{z^2}.
\end{equation}
Then the scalar curvature $R=-2$. In the classical theory if $Q=2\beta^{-1}\rho^{\vee}$
with the dual Weyl vector $\rho^{\vee}$ satisfying
\begin{equation}
\alpha_i \cdot \rho^{\vee}=1,~~~i=1,2,...,r,
\end{equation}
the equations of motion are invariant under the Weyl transformation
\begin{equation}
g_{ab}\rightarrow  e^{2\omega}g_{ab},~~~ \phi\rightarrow \phi-Q \omega.
\end{equation}
Therefore the theory is classically integrable because the flat space theory has a Lax pair.
The vector $Q$ needs to be modified at quantum level \cite{Mansfield:1982sq}.
In this paper we mainly focus on  classical aspects  of the theory so we simply set $Q=2\beta^{-1}\rho^{\vee}$.

The coefficients $n_i$ can be chosen arbitrarily through shifts in the scalar fields. If we require that the potential minimized at $\phi=0$, then  
\begin{equation}
\sum_i n_i \alpha_i= 2\rho^{\vee},~~~ \Rightarrow \sum_i n_i \alpha_i\cdot \alpha_j=2,~~~j=1,2,...,r.
\end{equation}
The  mass spectrum can be obtained by calculating the eigenvalues of the matrix $\sum_i n_i\alpha_i^a  \alpha_i^b$.
In the standard holographic dictionary, the mass of a scalar field $\phi_a$ can be expressed as $m^2_a=\Delta_{a}(\Delta_a-1)$ where  $\Delta_{a}$ is the conformal dimension of the boundary operator dual to $\phi_a$.
The spectrum of conformal dimensions   coincides with the set of  exponents plus one or equivalently degrees of fundamental adjoint-invariant polynomials of the corresponding Lie algebra.
The sets of these values are given in Table \ref{dim2}.
\begin{table}[H]
	\begin{center}
\begin{tabular}{ll}
	\hline 
Lie algebra	&  Conformal dimensions  \\ 
	\hline 
$a_n$	&  2, 3, ..., $n+1$ \\ 
	\hline 
$b_n$	&  2, 4, ..., $2n$ \\ 
	\hline 
$c_n$ 	&  2, 4, ..., $2n$\\ 
	\hline 
$d_n $	&  2, 4, ..., $2n-2$, $n$ \\ 
	\hline 
$e_6$ 	& 2, 5, 6, 8, 9, 12\\ 
	\hline 
$e_7$	& 2, 6, 8, 12, 14, 18 \\ 
	\hline 
$e_8$ 	& 2, 8, 12, 14, 18, 20, 24, 30\\ 
	\hline 
$f_4$	& 2, 6, 8, 12\\ 
\hline 
$g_4$	& 2, 6 \\ 
\hline 
\end{tabular} 
\end{center}
\caption{Spectrums of conformal dimensions in Toda field theories\label{dim2}}
\end{table}

Interestingly, the spins of generators in the $\mathcal{W}$-algebras are also equal to the exponents of the Lie algebras plus one, 
which implies that there is a one-to-one correspondence between the boundary operators and generators in the $\mathcal{W}$-algebras.
We will check this conjecture by computing the tree-level four-point correlation functions of boundary operators in the next section.

As some concrete examples, the potentials of Toda field theories with Lie algebras of rank 1 and 2  are given by 
\begin{align}
V_{a_1}=&\frac{2}{\beta^2}(e^{\beta\phi}-\beta\phi-1)=\phi^2+\frac{1}{3}\beta\phi^3+\frac{1}{12}\beta^2\phi^4+O\left(\beta ^3\right),\label{va1}\\
V_{a_2}=&\frac{2 }{\beta ^2}(e^{\beta  \phi _1} \cosh (\sqrt{3} \beta  \phi _2)-\beta  \phi _1-1) \label{va2}\nn\\
=&\left(\phi _1^2+3 \phi _2^2\right)+\beta  \left(\frac{\phi _1^3}{3}+3 \phi _2^2 \phi _1\right)+\beta ^2
\left(\frac{\phi _1^4}{12}+\frac{3}{2} \phi _2^2 \phi _1^2+\frac{3 \phi _2^4}{4}\right)+O\left(\beta ^3\right),\\
V_{b_2}=&\frac{1}{5	\beta ^2}(4 e^{\beta  \left(\phi _1+3 \phi _2\right)}+6 e^{\beta  \left(\phi _1-2 \phi _2\right)}-10 \beta  \phi _1-10)\label{vb2}\nn\\
=&\left(\phi _1^2+6 \phi _2^2\right)+\beta  \left(\frac{\phi _1^3}{3}+6 \phi _2^2 \phi _1+2 \phi _2^3\right)+\beta ^2
\left(\frac{\phi _1^4}{12}+3 \phi _2^2 \phi _1^2+2 \phi _2^3 \phi _1+\frac{7 \phi _2^4}{2}\right)+O\left(\beta
^3\right),
\\
V_{g_2}=&\frac{1}{7 \beta ^2}(-14 \beta  \phi _1+9 e^{\beta  \left(\phi _1-5 \phi _2/\sqrt{3}\right)}+5 e^{\beta  \left(\phi _1+3 \sqrt{3} \phi _2\right)})\label{vg2}\nn\\
=&\left(\phi _1^2+15 \phi _2^2\right)+\beta  \left(\frac{\phi _1^3}{3}+15 \phi _2^2 \phi _1+\frac{20 \phi _2^3}{\sqrt{3}}\right)\nn\\&
+\frac{1}{12} \beta ^2 \left(\phi _1^4+90 \phi _2^2 \phi _1^2+80 \sqrt{3} \phi _2^3 \phi _1+305 \phi _2^4\right)+O\left(\beta ^3\right).
\end{align}

\section{Four-point functions in Toda field theories}\label{s3}

\subsection{Witten diagrams in $AdS_2$}
We now briefly review  some basic ingredients of Witten diagrams in $AdS_2$.
 Witten diagrams are Feynman diagrams built from
 bulk-to-bulk propagators and  bulk-to-boundary propagators.
 The  bulk-to-bulk propagator of a scalar with mass $m^2=\Delta(\Delta-1)$  is given by
\begin{equation}
G_\Delta(x_1,z_1,x_2,z_2)=C_{\Delta} u^{-\Delta} {}_2F_{1}(\Delta,\frac{2\Delta}{2},2\Delta ,-\frac{4}{u}),
\end{equation}
where
\begin{equation}
u=\frac{(x_1-x_2)^2+(z_1-z_2)^2}{z_1 z_2},~~~C_{\Delta}=\frac{\Gamma(\Delta)}{2 \sqrt{\pi} \Gamma(\Delta+1/2)}.
\end{equation}
The  bulk-to-boundary propagator is given by
\begin{equation}
K_\Delta(x,x_0,z_0)=C_{\Delta} \left(\frac{z_0}{z_0^2+(x-x_0)^2}\right)^\Delta.
\end{equation}
To get correlation functions for normalized operators whose  two-point functions have unit coefficient, every  external line has to be multiplied by a factor $C_\Delta^{-1/2}$. For instance, an $n$-point contact diagram takes the form
\begin{equation}
W^{\mathrm{contact}}_{\Delta_i}(x_i)=\int \frac{d x_0 d z_0}{z_0^2}  \prod_{i=1}^n  C_{\Delta_i}^{-1/2}  K_{\Delta_i}(x_i,x_1,z_1).
\end{equation}
where $\Delta_i$, $i=1,...,n$, are conformal dimensions of the external operators. The expression for $n=3$ can be found in \cite{Freedman:1998tz}
\begin{equation}\label{3pt}
\begin{split}
&W^{\mathrm{contact}}_{\Delta_1\Delta_2\Delta_3}(x_1,x_2,x_3)\\
=&\prod_{i=1}^3  C_{\Delta_i}^{1/2}
\frac{\sqrt{\pi } \Gamma (\frac{\Delta _1+\Delta _2-\Delta _3}{2}) 
	\Gamma (\frac{\Delta _1+\Delta _3-\Delta _2}{2})
	\Gamma (\frac{\Delta _2+\Delta _3-\Delta _1}{2})	 
	\Gamma (\frac{\Delta _1+\Delta _2+\Delta_3-1}{2})}{
	2 \Gamma \left(\Delta _1\right) \Gamma \left(\Delta _2\right) \Gamma \left(\Delta _3\right)
	|x_{12}|^{\Delta _1+\Delta _2-\Delta _3}
	|x_{23}|^{\Delta _2+\Delta _3-\Delta _1}
	|x_{13}|^{\Delta _1+\Delta _3-\Delta _2}},
\end{split}
\end{equation}
where $x_{ij}\equiv x_i-x_j$.

We shall concentrate on tree-level four-point functions in Toda field theories. 
The contact Witten diagrams generated by quartic interactions
 can be written in terms of the $D$-functions \cite{Liu:1998ty,DHoker:1999kzh,Dolan:2003hv} defined as
\begin{equation}
D_{\Delta_1\Delta_2\Delta_3\Delta_4}(x_i)=\int \frac{d x_0 d z_0}{z_0^2}  \prod_{i=1}^4 \left(\frac{z_0}{z_0^2+(x_i-x_0)^2}\right)^{\Delta_i}.
\end{equation}
It is convenient to work with the reduced $\bar D$-functions which are functions of the cross-ratio  (not to confuse with the same symbol used for $AdS_2$ coordinate)
\begin{equation}
z=\frac{x_{12}x_{34}}{x_{13}x_{24}}.
\end{equation}
They are defined by extracting a kinematic factor
\begin{equation}
D_{\Delta_1\Delta_2\Delta_3\Delta_4}(x_i)= \frac{\sqrt{\pi} \Gamma(\Sigma -1/2)}{2 \prod_{i=1}^4 \Gamma(\Delta_i) } 
\frac{x_{14}^{2(\Sigma-\Delta_1-\Delta_4)}x_{34}^{2(\Sigma-\Delta_3-\Delta_4)}}{x_{13}^{2(\Sigma-\Delta_2)}x_{24}^{2\Delta_2}}
\bar D_{\Delta_1\Delta_2\Delta_3\Delta_4}(z),
\end{equation}
with $\Sigma=\sum_{i=1}^4 \Delta_i/2$. $\bar D$-functions with integer indices can be evaluated recursively by using the identities in \cite{Arutyunov:2002fh}.
 We list some of the $\bar D$-functions in Appendix \ref{app1}.

We also need to consider exchange Witten diagrams mediated by the cubic interactions. An $s$-channel exchange Witten diagram can be written as
\begin{equation}
\begin{split}
W^{s}_{\Delta_1\Delta_2\Delta_3\Delta_4,\Delta_E} (x_i)=&\prod_{i=1}^4  C_{\Delta_i}^{-1/2}\int \frac{d y_1 d w_1}{w_1^2}\frac{d y_2 d w_2}{w_2^2}   
 K_{\Delta_1}(x_1,y_1,z_1)K_{\Delta_2}(x_2,y_1,w_1)\\
 &\times G_{\Delta_E}(y_1,w_1,y_2,w_2)K_{\Delta_3}(x_3,y_2,w_2)K_{\Delta_4}(x_4,y_2,w_2).
\end{split}
\end{equation}
where $\Delta _E$ is the conformal dimension of the exchange operator. 
In the special case when $k=(\Delta_1+\Delta_2-\Delta_E)/2$ is a positive integer, this diagram can be written as a sum of $D$-functions
\cite{DHoker:1999mqo}
\begin{equation}\label{wtod}
W^{s}_{\Delta_1\Delta_2\Delta_3\Delta_4,\Delta_E} (x_i)=\prod_{i=1}^4  C_{\Delta_i}^{1/2} \sum_{l=1}^{k} 
\frac{(\Delta_1)_{-l} (\Delta_2)_{-l}D_{\Delta_1-l\,\Delta_2-l\,\Delta_3\Delta_4}(x_i)}{4(\frac{\Delta_1+\Delta_2-\Delta_E}{2})_{1-l} (\frac{\Delta_1+\Delta_2-1+\Delta_E}{2})_{1-l}x_{12}^{2l}  } .
\end{equation}
where the standard Pochhammer symbol $(a)_b=\Gamma(a+b)/\Gamma(a)$ is used.

\subsection{$a_1$ theory}

We start with the Liouville ($a_1$ Toda) field theory. The potential is given in (\ref{va2}).
The scalar field with $m^2=2$ corresponds to a boundary operator with conformal dimension $\Delta=2$.
The tree-level four-point function receives contributions from a quartic contact diagram and  exchange diagrams in $s$-, $t$- and $u$-channels.
We have
\begin{equation}
\begin{split}
\langle O(x_1)O(x_2)O(x_3)O(x_4)\rangle_{\mathrm{tree}}
=&\beta^2(4W^{s}_{2222,2}+4W^{t}_{2222,2}+4W_{2222,2}^{u}-2 W_{2222}^{\mathrm{contact}})\\
=&\frac{4\beta^2}{9\pi^2}(x_{34}^{-2} D_{2211}+x_{14}^{-2} D_{1221}+x_{24}^{-2} D_{2121}-2 D_{2222})\\
=&\frac{\beta^2}{6\pi x_{13}^{4}x_{24}^{4}}(\bar D_{2211}+\bar D_{1221}+\bar D_{2121}-5 \bar D_{2222}),
\end{split}
\end{equation}
where we have used (\ref{wtod}) to write exchange diagrams in terms of $D$-functions.
Using the explicit  expressions for the $\bar D$-functions  given in Appendix \ref{app1}, we get
\begin{equation}\label{c2222}
\begin{split}
&\bar D_{2211}+\bar D_{1221}+\bar D_{2121}-5 \bar D_{2222}\\
=&\frac{z^3 \log (\left| z\right| )+\left(-z^3+3 z-2\right) \log (\left| 1-z\right| )-z^2+z}{3 (z-1)^2 z^3}\\
&+\frac{(3-z) z^2 \log (\left| z\right| )+(z-1)^3 \log (\left| 1-z\right| )+z^2-z}{3 (z-1)^3 z^2}\\
&+\frac{z^2 (2 z-3) \log (\left| z\right| )-(2 z+1) (z-1)^2 \log (\left| 1-z\right| )+z^2-z}{3 (z-1)^2 z^2}\\
&+\frac{z^3 \left(-2 z^2+5 z-5\right) \log (\left| z\right| )+\left(2 z^2+z+2\right) (z-1)^3 \log (\left| 1-z\right| )+2 z \left(z^3-2 z^2+2 z-1\right)}{3 (z-1)^3 z^3}\\
=& \frac{z^2-z+1}{(z-1)^2 z^2}.
\end{split}
\end{equation}
and thus
\begin{equation}\label{c2222}
\langle O(x_1)O(x_2)O(x_3)O(x_4)\rangle_{\mathrm{tree}}=
\frac{\beta^2}{12\pi}\Big(
\frac{1}{x_{12}^2x_{23}^2x_{34}^2x_{14}^2}+
\frac{1}{x_{13}^2x_{23}^2x_{24}^2x_{14}^2}+
\frac{1}{x_{12}^2x_{24}^2x_{34}^2x_{13}^2}
\Big).
\end{equation}

 This looks like the connected part of a four-point function  of the stress-tensor operators $T$ in the Virasoro algebra up to an overall coefficient, 
 although it is a one-dimensional correlation function. Furthermore, the disconnected parts of both four-point functions also take the same form.

Taking into account carefully the overall coefficient,
the central charge $c$ of  Virasoro algebra is related to the $\beta$ by $c = 48\pi \beta^{-2}$ and $O(x)$ corresponds to $-\sqrt{2/c}T(w)$.
This is consistent with tree-level three-point function computing by (\ref{3pt})
\begin{equation}
\langle O(x_1)O(x_2)O(x_3)\rangle_{\mathrm{tree}}
	=-\frac{\beta }{\sqrt{6 \pi }} \frac{1}{x_{12}^2x_{23}^2x_{13}^2}
\sim-\sqrt{\frac{8}{c^3}}\langle T(w_1)T(w_2)T(w_3)\rangle.
\end{equation}

The central charge of Liouville theory in our convention is
\begin{equation}
c_{a_1}=1+ 48 \pi  \left(\frac{\beta }{8 \pi }+\frac{1}{\beta }\right)^2
\end{equation}
In the classical limit $c_{a_1}\rightarrow  48\pi \beta^{-2}$. 
One can expect that at the quantum level the central charge $c$ of  Virasoro algebra is equal to the central charge of Liouville theory.
If this conjecture is true,  $n$-point functions beyond tree-level can be obtained from correlation functions of the stress-tensor operators $T(w_i)$.

\subsection{$a_2$ theory}
We now  turn to the $a_2$ theory. 
The boundary operators have conformal dimensions $\Delta_1=2$ and $\Delta_2=3$, which are equal to the spins of generators in the
 algebra $\mathcal{W}(2,3)$. The notation $\mathcal{W}(2,s_1,s_2,...)$ means that the algebra is generated by
primary currents $W^{(s_i)}$ of spins $s_i$ together with the Virasoro generator $T$ of spin 2.  

The  four-point function of four $O_1$ is the same as (\ref{c2222}). The other independent four-point functions are
\begin{equation}
\begin{split}\label{c2233}
&\langle O_1(x_1)O_1(x_2)O_2(x_3)O_2(x_4)\rangle_{\mathrm{tree}}\\
=&\beta^2(12W^{s}_{2233,2}+36W^{t}_{2233,3}+36W_{2233,3}^{u}-6 W^{\mathrm{contact}}_{2233})\\
=&\frac{16\beta^2}{45\pi^2}(3x_{12}^{-2} D_{1133}+\frac{9}{2}x_{14}^{-2} D_{1232}+\frac{9}{2}x_{24}^{-2} D_{2132}-6 D_{2233})\\
=&\frac{\beta^2}{4 \pi  x_{13}^4 x_{24}^4 x_{34}^2}
(z^{-2}\bar D_{1133}+3\bar D_{1232}+3\bar D_{2132}-7 \bar D_{2222})\\
=&\frac{\beta^2}{8\pi {  x_{13}^4 x_{24}^4 x_{34}^2}} \frac{3 z^2-2 z+2}{  (z-1)^2 z^2}\\
=&\frac{\beta^2}{8\pi}\Big(
\frac{1}{x_{12}^2x_{23}^2x_{34}^4x_{14}^2}+\frac{1}{x_{12}^2x_{24}^2x_{34}^4x_{13}^2}+
\frac{2}{x_{13}^2x_{23}^2x_{24}^2x_{14}^2x_{34}^2}
\Big),
\end{split}
\end{equation}
and
\begin{equation}
\begin{split}\label{c3333}
&\langle O_2(x_1)O_2(x_2)O_2(x_3)O_2(x_4)\rangle_{\mathrm{tree}}\\
=&\beta^2(36W^{s}_{3333,2}+36W^{t}_{3333,2}+36W_{3333,2}^{u}-18 W^{\mathrm{contact}}_{3333})\\
=&\frac{64\beta^2}{225\pi^2}(\frac{45}{8}(x_{34}^{-4} D_{3311}+x_{14}^{-4} D_{1331}+x_{24}^{-4} D_{3131})
+\frac{9}{4}(x_{34}^{-2} D_{3322}+x_{14}^{-2} D_{2332}\\&+x_{24}^{-2} D_{3232})-18 D_{3333})\\
=&\frac{3\beta^2}{40\pi x_{13}^6 x_{24}^6}(5(\bar D_{3311}+\bar D_{1331}+\bar D_{3131})
+7(\bar D_{3322}+\bar D_{2332}+\bar D_{3232})-63 \bar D_{3333})\\
=&\frac{\beta^2}{x_{13}^6 x_{24}^6}\frac{3 \left(10 z^6-30 z^5+29 z^4-8 z^3+29 z^2-30 z+10\right)}{80 \pi  (z-1)^4 z^4}\\
=&\frac{3\beta^2}{80\pi}\Big(
\frac{5}{x_{12}^4x_{23}^2x_{34}^4x_{14}^2}+
\frac{5}{x_{13}^4x_{23}^2x_{24}^4x_{14}^2}+
\frac{5}{x_{12}^4x_{24}^2x_{34}^4x_{13}^2}+
\frac{5}{x_{12}^2x_{23}^4x_{34}^2x_{14}^4}\\&+
\frac{5}{x_{13}^2x_{23}^4x_{24}^2x_{14}^4}+
\frac{5}{x_{12}^2x_{24}^4x_{34}^2x_{13}^4}-\frac{16}{x_{12}^2x_{24}^2x_{34}^2x_{13}^2x_{14}^2x_{23}^2}
\Big).
\end{split}
\end{equation}

Comparing them to the four-point functions in the algebra $\mathcal{W}(2,3)$ \cite{Zamolodchikov:1985wn}
\begin{align}
&\langle T(w_1)T(w_2)W^{(3)}(w_3)W^{(3)}(w_4)\rangle\nn\\
=&{\frac{1} {w_{13}^4 w_{24}^4 w_{34}^2}}\left(\frac{ c^2}{6z^2}+c\left(\frac{2}{z^2}+\frac{3}{(1-z)^2}+\frac{2}{z}+\frac{2}{1-z}\right)\right),\label{w2233}\\[12pt]
&\langle W^{(3)}(w_1)W^{(3)}(w_2)W^{(3)}(w_3)W^{(3)}(w_4)\rangle\nn\\
=&\frac{1}{w_{13}^6 w_{24}^6}\bigg(  \frac{c^2}{9} \left(\frac{1}{z^6}+\frac{1}{(1-z)^6}+1\right)\nn\\&
 +c \left(\frac{2}{z^4}+\frac{2}{(1-z)^4}
 +\frac{2}{z^3}+\frac{2}{(1-z)^3}+\frac{9}{5 z^2}+\frac{9}{5 (1-z)^2}+\frac{8}{5 z}+\frac{8}{5 (1-z)}\right)\nn\\&
+\frac{32}{5} \frac{16 c}{5 c+22} \left(\frac{1}{z^2}+\frac{1}{(1-z)^2}+\frac{2}{z}+\frac{2}{1-z}\right) \bigg),\label{w3333}
\end{align}
one find the tree-level four-point functions (\ref{c2233}) and (\ref{c3333}) correspond to  the order $c$ contributions to (\ref{w2233}) and (\ref{w3333}) respectively.
The disconnected pieces of the four-point functions correspond to the order $c^2$ contributions.
Therefore, we get the expected correspondence.
This  correspondence provides a useful way to obtain higher-point functions, because correlation functions of $\mathcal{W}$-currents and stress-tensor $T$  can be computed by using recursion relations
derived  in \cite{Zamolodchikov:1985wn}.

\subsection{$b_2$ theory}
Finally we consider the $b_2$ theory. The conformal dimensions of the boundary operators are $\Delta_1=2$ and $\Delta_2=4$. 
The  tree-level four-point functions are
\begin{align}
&\langle O_1(x_1)O_1(x_2)O_2(x_3)O_2(x_4)\rangle_{\mathrm{tree}}\nn\\
=&\beta^2(24W^{s}_{2244,2}+144W^{t}_{2244,4}+144W_{2244,4}^{u}-12 W^{\mathrm{contact}}_{2244})\nn\\
=&\frac{\beta^2}{6 \pi  x_{13}^4 x_{24}^4 x_{34}^4}
(z^{-2}\bar D_{1144}+6\bar D_{1243}+6\bar D_{1234}-3 \bar D_{2244})\nn\\
=&\frac{\beta^2}{3\pi {  x_{13}^4 x_{24}^4 x_{34}^4}}\frac{2 z^2-z+1}{ (z-1)^2 z^2},\\[12pt]
&\langle O_1(x_1)O_2(x_2)O_2(x_3)O_2(x_4)\rangle_{\mathrm{tree}}\nn\\
=&\beta^2(144W^{s}_{2444,4}+144W^{t}_{2444,4}+144W_{2444,4}^{u}-12 W^{\mathrm{contact}}_{2444})\nn\\
=&\sqrt{\frac{3}{70}} \frac{\beta^2 x_{14}^2}{\pi x_{13}^6 x_{24}^8 x_{34}^2} 
\big(3(z^{-2}\bar D_{1344}+\bar D_{1443}+\bar D_{1434})
-\frac{11}{2} \bar D_{2444}\big)\nn\\
=&\sqrt{\frac{3}{70}} \frac{\beta^2 x_{14}^2}{\pi x_{13}^6 x_{24}^8 x_{34}^2} \frac{z^2-z+1}{(z-1)^4 z^2},\\[12pt]
&\langle O_2(x_1)O_2(x_2)O_2(x_3)O_2(x_4)\rangle_{\mathrm{tree}}\nn\\
=&\beta^2(144(W^{s}_{4444,4}+W^{t}_{4444,4}+W_{4444,4}^{u}
+W^{s}_{4444,2}+W^{t}_{4444,2}+W_{4444,2}^{u})
-84 W^{\mathrm{contact}}_{4444})\nn\\
=&
\frac{\beta^2}{\pi x_{13}^8 x_{24}^8}\big(\frac{1}{3}(\bar D_{4411}+\bar D_{1441}+\bar D_{4141})
+\frac{69}{70}(\bar D_{4422}+\bar D_{2442}+\bar D_{4242})\nn\\&
+\frac{33}{35}(\bar D_{4433}+\bar D_{3443}+\bar D_{4343})
-\frac{143}{20} \bar D_{4444}\big)\nn\\
=&\frac{\beta^2}{420 \pi x_{13}^8 x_{24}^8} 
\frac{\left(z^2-z+1\right)^2 \left(280 z^6-840 z^5+279 z^4+842 z^3+279 z^2-840 z+280\right)}{  (z-1)^6 z^6},\label{c4444}
\end{align}
together with $\langle O_1(x_1)O_1(x_2)O_1(x_3)O_1(x_4)\rangle_{\mathrm{tree}}$ given by (\ref{c2222}).

As an application of the conjectured correspondence, we now demonstrate how to compute (\ref{c4444}) using recursion relations.
The operator product expansion of two spin-4 currents $W^{(4)}$ in the algebra $\mathcal{W}(2,4)$ can be found in 
\cite{Bouwknegt:1988sv,Hamada:1988vw,Zhang:1989iz,Blumenhagen:1990jv,Kausch:1990bn}, and in the large central charge limit we have
\begin{equation}
\begin{split}
W^{(4)}(z)W^{(4)}(w)=&\frac{c/4}{(z-w)^8}+\sum_{k=0}^{5}\frac {12 (k + 1)} {\Gamma (k + 4)} \frac{\partial^k T(w)}{(z-w)^{6-k}}
\\
&+\sum_{k=0}^{3}\frac{72 \sqrt{105} (k+1)_3}{\Gamma (k+8)}\frac{\partial^k W^{(4)}(w)}{(z-w)^{4-k}}+O\left(c^{-1}\right).
\end{split}
\end{equation}
Following similar analysis in \cite{Zamolodchikov:1985wn}, we have the recursion relation
\begin{equation}
\begin{split}
&\langle W^{(4)}(w_1)W^{(4)}(w_2)...W^{(4)}(w_n)\rangle\\
=&
\sum_{i=2}^{n}\frac{c/4}{(w_1-w_2)^8}\langle W^{(4)}(w_2)...W^{(4)}(w_{i-1})W^{(4)}(w_{i+1})...W^{(4)}(w_n)\rangle\\
&
+\sum_{i=2}^{n}\sum_{k=0}^{5}\frac {12 (k + 1)} {\Gamma (k + 4)} \frac{\partial^k_{w_i} }{(w_1-w_i)^{6-k}}
\langle W^{(4)}(w_2)...W^{(4)}(w_{i-1})T(w_i)W^{(4)}(w_{i+1})...W^{(4)}(w_n)\rangle
\\
&+\sum_{i=2}^{n}\sum_{k=0}^{3}\frac{72 \sqrt{105} (k+1)_3}{\Gamma (k+8)}\frac{\partial^k_{w_i}}{(w_1-w_i)^{4-k}}\langle W^{(4)}(w_2)...W^{(4)}(w_n)\rangle
+O\left(c^{n/2-2}\right).
\end{split}
\end{equation}
Using the correspondence $O_1\sim -\sqrt{2/c}T$ and $O_2\sim -\sqrt{4/c}W^{(4)}$, we get
\begin{equation}
\begin{split}\label{r4444}
&\langle O_2(x_1)O_2(x_2)O_2(x_3)O_2(x_4)\rangle_{\mathrm{tree}}\\
=
&
-\sum_{k=0}^{5}\frac{\beta }{\sqrt{6 \pi }}\frac {12 (k + 1)} {\Gamma (k + 4)}
\Big( \frac{\partial^k_{w_2} }{(w_1-w_2)^{6-k}}
\langle O_1(x_2)O_2(x_3)O_2(x_4)\rangle_{\mathrm{tree}}
\\
&+\frac{\partial^k_{w_3} }{(w_1-w_3)^{6-k}}
\langle O_2(x_2)O_1(x_3)O_2(x_4)\rangle_{\mathrm{tree}}+\frac{\partial^k_{w_4} }{(w_1-w_4)^{6-k}}
\langle O_2(x_2)O_2(x_3)O_1(x_4)\rangle_{\mathrm{tree}}\Big)
\\
&-\sum_{i=2}^{4}\sum_{k=0}^{3} \frac{36 \sqrt{35}\beta}{ \sqrt{ \pi }}\frac{ (k+1)_3}{\Gamma (k+8)}\frac{\partial^k_{w_i}}{(w_1-w_i)^{4-k}}
\langle O_2(x_2)O_2(x_3)O_2(x_4)\rangle_{\mathrm{tree}}.
\end{split}
\end{equation}
The three-point functions can be obtained by use of (\ref{3pt})
\begin{align}
&\langle O_1(x_2)O_2(x_3)O_2(x_4)\rangle_{\mathrm{tree}}=-\frac{2\beta}{\sqrt{6 \pi }} \frac{1}{ {  x_{23}^2 x_{24}^2 x_{34}^6}},\\
&\langle O_2(x_2)O_2(x_3)O_2(x_4)\rangle_{\mathrm{tree}}=-\frac{3 \beta }{2 \sqrt{35 \pi }}\frac{1}{ {  x_{23}^4 x_{24}^4 x_{34}^4}}.
\end{align}
Plugging them into (\ref{r4444}) one immediately obtain the expected result (\ref{c4444}).
It is reasonable to expect that higher-point functions can be obtained in the same way.

\section{Conclusions}\label{sc}
In this paper, we have studied the	boundary correlation functions in Toda field theories in $AdS_2$.
We have found a relation between boundary operators and generators  in the $\mathcal{W}$-algebra
 by computing  tree-level four-point functions in some simple Toda field theories.
The  conformal dimensions of boundary operators are equal to the spins of generators in the $\mathcal{W}$-algebra, 
and the tree-level four-point functions correspond to the
 connected four-point functions of the $\mathcal{W}$-currents or the stress-tensor in the large central charge.
We  conjecture the correspondence holds at the quantum level.
If rigorously proved, this conjecture will provide useful information about higher-loop and higher-point Witten diagrams in 
$AdS_2$.
It would be interesting to extend our study to the supersymmetric  Toda field theories.

\section*{Acknowledgments}
I would like to thank Konstantin Zarembo for motivating me to  work on the topic of this paper and making valuable suggestions.
I also thank Zhibin Li and Jun-Bao Wu for helpful discussions.
This work was supported by the grant ``Exact Results in Gauge and String Theories'' from the Knut and Alice Wallenberg foundation.

\appendix

\section{$\bar D$-functions}\label{app1}
There are  two independent cross-ratios $u$ and $v$  in higher dimensions.
$\bar D$-functions in one dimension can be obtained from the higher dimensional ones by
\begin{equation}
\bar D_{\Delta_1\Delta_2\Delta_3\Delta_4}(z)=\lim_{\substack{
	u \rightarrow z^2\\v\rightarrow (1-z)^2}} \bar D_{\Delta_1\Delta_2\Delta_3\Delta_4}(u,v),
\end{equation}
where $\bar D_{\Delta_1\Delta_2\Delta_3\Delta_4}(u,v)$ are $\bar D$-functions in general dimensions which can be computed recursively starting from
\begin{equation}
 \bar D_{1111}(u,v)=\Phi(u,v).
\end{equation} 
The standard four-dimensional one-loop integral $\Phi(u,v)$ can be found in \cite{Usyukina:1992jd} and the  recursion relations can be found in \cite{Arutyunov:2002fh}.
After taking the  limit $u \rightarrow z^2$ and $v\rightarrow (1-z)^2$, we have

\begin{align}
\bar D_{1111}=&\frac{2 \log (\left| z\right| )}{z-1}-\frac{2 \log (\left| 1-z\right| )}{z},\\
\bar D_{2211}=&\frac{z^3 \log (\left| z\right| )+\left(-z^3+3 z-2\right) \log (\left| 1-z\right| )-z^2+z}{3 (z-1)^2 z^3},\\
\bar D_{1221}=&\frac{(3-z) z^2 \log (\left| z\right| )+(z-1)^3 \log (\left| 1-z\right| )+z^2-z}{3 (z-1)^3 z^2},\\
\bar D_{2121}=&\frac{z^2 (2 z-3) \log (\left| z\right| )-(2 z+1) (z-1)^2 \log (\left| 1-z\right| )+z^2-z}{3 (z-1)^2 z^2},\\
\bar D_{2222}=&\frac{1}{15 (z-1)^3 z^3}
\Big(
-(z-1)^3 \left(2 z^2+z+2\right) \log (\left| 1-z\right| )-2 z^4+4 z^3-4 z^2+2 z
\nn\\&
+z^3 \left(2 z^2-5 z+5\right) \log (\left| z\right| )
\Big),
\end{align}
{\small
\begin{align}
\bar D_{1232}=&\frac{1}{15 (z-1)^4 z^2}
\Big(
(2 z+3) (z-1)^4 \log (\left| 1-z\right| )+2 z^4-4 z^3-z^2+3 z
-z^4 (2 z-5) \log (\left| z\right| )
\Big),\\
\bar D_{2132}=&\frac{(5-3 z) z^4 (\log (\left| 2-2 z\right| )-\log (2 \left| z\right| ))+(3-5 z) \log (\left| 1-z\right| )+(1-z) (z (3 z-8)+3) z}{15 (z-1)^3 z^2},
\end{align}
\begin{align}
\bar D_{3311}=&\frac{1}{15 (z-1)^3 z^5}
\Big(
-2 (z-1)^3 \left(z^2+3 z+6\right) \log (\left| 1-z\right| )-2 z^4-z^3+6 z^2-3 z
\nn\\&
+2 z^5 \log (\left| z\right| )
\Big),\\
\bar D_{3322}=&\frac{1}{210	(z-1)^4 z^5}
\Big(-18 z^6+47 z^5-34 z^4-43 z^3+72 z^2-24 z\nn\\&
+2 z^5 \left(9 z^2-28 z+28\right) \log (\left| z\right| )-2 (z-1)^4 \left(9 z^3+8 z^2+6 z+12\right) \log (\left| 1-z\right| )
\Big),\\
\bar D_{3333}=&-\frac{2}{315 (z-1)^5 z^5}
\Big(12 z^8-48 z^7+73 z^6-51 z^5+51 z^4-73 z^3+48 z^2+12 z\nn\\&
+6 (z-1)^5 \left(2 z^4+z^3+z^2+z+2\right) \log (\left| 1-z\right| )\nn\\&
-6 z^5 \left(2 z^4-9 z^3+16 z^2-14 z+7\right) \log (\left| z\right| )\Big),
\end{align}

\begin{align}
\bar D_{2134}=&\frac{1}{70 (z-1)^3 z^2}\Big(
6 z^6-11 z^5-5 z^4-9 z^3+39 z^2-20 z-2 z^6 (3 z-7) \log (\left| z\right|) \nn\\&
+2 (z-1)^5 \left(3 z^2+8 z+10\right) \log (\left| 1-z\right| )\Big),\\
\bar D_{1344}=&\frac{1}{315 (z-1)^6 z^3}\Big(30 z^8-93 z^7+64 z^6+183 z^5-426 z^4+422 z^3-240 z^2+60 z
 \nn\\&
+6 (z-1)^6 \left(5 z^3+12 z^2+15 z+10\right) \log (\left| 1-z\right| )-6 z^7 \left(5 z^2-18 z+18\right) \log (\left| z\right| )
\Big),\\
\bar D_{2444}=&\frac{1}{1155 (-1 + z)^7 z^5}\Big(
-12 z^7 \left(10 z^4-55 z^3+121 z^2-132 z+66\right) \log (\left| z\right| )
\nn\\&
+12 (z-1)^7 \left(10 z^4+15 z^3+16 z^2+15 z+10\right) \log (\left| 1-z\right| )\nn\\&
+2 \left(60 z^{10}-300 z^9+581 z^8-524 z^7+287 z^6-287 z^5+524 z^4-581 z^3+300 z^2-60 z\right)
\Big),
\end{align}
\begin{align}
\bar D_{4411}=&\frac{z^3 (z (2 z-5)+5) \log (\left| z\right| )-\left(2 z^2+z+2\right) (z-1)^3
	\log (\left| 1-z\right| )-2 z ((z-1) z+1) (z-1)}{15 (z-1)^3 z^3},\\
\bar D_{4422}=&\frac{1}{630 (-1 + z)^5 z^7}\Big(
-48 z^8+138 z^7-97 z^6-39 z^5-231 z^4+637 z^3-480 z^2+120 z
\nn\\&
-6 (z-1)^5 \left(8 z^4+13 z^3+12 z^2+10 z+20\right) \log (\left| 1-z\right| )\nn\\&
+6 z^7 \left(8 z^2-27 z+27\right) \log (\left| z\right| )
\Big),\\
\bar D_{4433}=&\frac{1}{3465 (z-1)^6 z^7}\Big(
6 z^7 \left(50 z^4-264 z^3+561 z^2-594 z+297\right) \log (\left| z\right| )
-2 (150 z^{10}-717 z^9\nn\\&
+1337 z^8-1167 z^7+393 z^6+751 z^5-1836 z^4+1809 z^3-900 z^2+180 z)\nn\\&
-6 (z-1)^6 \left(50 z^5+36 z^4+27 z^3+28 z^2+30 z+60\right) \log (\left| 1-z\right| )
\Big),\\
\bar D_{4444}=&\frac{1}{5005 (-1 + z)^7 z^7}\Big(
6 z^7 \left(100 z^6-650 z^5+1794 z^4-2717 z^3+2431 z^2-1287 z+429\right) \log (\left| z\right| )
\nn\\&
-2 z \left(z^2-z+1\right)^2 \left(300 z^7-1200 z^6+1207 z^5+579 z^4-579 z^3-1207 z^2+1200 z-300\right)\nn\\&
-6 (z-1)^7 \left(100 z^6+50 z^5+44 z^4+41 z^3+44 z^2+50 z+100\right) \log (\left| 1-z\right| )
\Big).
\end{align}
}

There are crossing relations
\begin{equation}
\begin{split}
\bar D_{\Delta_1\Delta_2\Delta_3\Delta_4}(z)=\bar D_{\Delta_3\Delta_2\Delta_1\Delta_4}(1-z)=z^{-\Delta_1-\Delta_2-\Delta_3+\Delta_4}\bar D_{\Delta_1\Delta_3\Delta_2\Delta_4}(1/z).\\
\end{split}
\end{equation}


\providecommand{\href}[2]{#2}\begingroup\raggedright\endgroup

\end{document}